# Networkcontrology

Adilson E. Motter

Department of Physics and Astronomy, Northwestern University, Evanston, IL 60208, USA

Northwestern Institute on Complex Systems (NICO), Northwestern University, Evanston, IL 60208, USA

**Abstract:** An increasing number of complex systems are now modeled as networks of coupled dynamical entities. Nonlinearity and high-dimensionality are hallmarks of the dynamics of such networks but have generally been regarded as obstacles to control. Here I discuss recent advances on mathematical and computational approaches to control high-dimensional nonlinear network dynamics under general constraints on the admissible interventions. I also discuss the potential of network control to address pressing scientific problems in various disciplines.

**Imagine a world without power grids, the Internet, transportation infrastructure, banking systems; a world without social structure, ecosystems, biogeochemical cycles, without life. These are just some of the many features that would be missing had the world been devoid of networks. Perhaps not surprisingly, networks have long been part of our scientific literacy—according to the Thomson Reuters database, no less than 0.5 million (out of 60 million) scientific papers published since 1900 have the word "network" in the title (not to mention elsewhere in the paper, or related terms such as web and graph). But what is the most substantial research question that can be formulated in the current study of networks? I would argue it is "how to rationally control the dynamics of complex networks." Why? Because the solution to this problem would address some of the most pressing biomedical, engineering, and socioeconomic questions of our time—from cell reprogramming, drug target identification and microbial strain optimization, to the development of smart self-healing infrastructure systems and of sustainable management of ecosystems, to the mitigation of cascading failures and contagion dynamics in financial, social and technological systems. Here I discuss why in many systems our current ability to control such networks is limited not by the available technology to actuate specific network elements, but instead by the challenges that nonlinearity, high dimensionality, and constraints on the interventions impose on designing system-level control actions. And I discuss promising new approaches to overcome these challenges and even benefit from the network properties that give rise to them.**

## I. INTRODUCTION

We have come a long way since the time researchers believed they were on the brink of discovering everything they needed to know about the physical world, as famously expressed by Laplace [1]. Whether because of quantum mechanics, chaos, or the fact that we don't even know what most of the universe is made of, now, 200 years later, very few researchers would embrace that optimistic view. Yet, a surprising number of researchers still reason along the following lines: if we could understand all the physics, they would say, then we could make sense of all the chemistry; and if we could understand all the chemistry, we could understand all the biology. After all, the argument goes, chemistry is applied quantum mechanics, and according to this simplified view, organelles are bags of chemicals, so cells are bags of bags of chemicals, and so on. But this program too seems hard to be completed, because as we go up in this hierarchy the systems become increasingly more structured, and it's no longer sufficient to understand the properties of the component parts in order to understand the properties of the system. That is to say, the systems become increasingly more complex.

But what exactly is a complex system? I will define it as a system that 1) is made up of a large number of interacting component parts and 2) exhibits collective dynamical behavior that cannot be anticipated from the properties of the parts themselves, not even in principle. For example, graphene, graphite, diamond, and all other allotropes of carbon are made of the same component parts—carbon atoms. Yet, they have very different physical properties; the key difference is in the network of interactions between those atoms. That is, in complex systems the interactions can be just as important as the parts themselves, or more, in determining the resulting properties. Thus, such systems lend themselves naturally to be modeled as networks of interactions between the component parts—a realization that is at the base of the modern, interdisciplinary study of networks initiated in the late 1990's [2, 3]. The question of tracing a cell or even a molecule to the parts they are composed of is therefore akin to the question of understanding the underlying networks, a problem that is as mathematical as it is physical in nature. It follows that the control of such systems can in principle be based on manipulating either their dynamical units or the interactions between them (or both).

That complex systems require different approaches and lead to different behavior has long been appreciated in physics. It is the basis of condensed matter physics and was popularized P. Anderson in his classic "More is Different" piece [4]. A key departure from that research is that, while Anderson could illustrate his ideas using the ammonia molecule, a system composed of four atoms, the current network-based study of complexity focuses on systems like metabolic networks, which in typical cells consist of an irreducible set of thousands of coupled biochemical reactions and a comparable (even if smaller) number of chemical species [5].[1] The latter is a network of interest to us in the context of control both because of applications to the optimizations of microbial organisms for the production of chemicals of industrial interest and because of applications to the development of therapeutic interventions for humans and other organisms.

Going from chemical reactions to living cells, another complex system of current interest in this context is that of neuronal networks. Consisting of tens of billions of dynamical units coupled together by some ten thousand times that number of connections, the brain is an example par excellence of a system whose functioning relies on its interconnectivity; and the processing capacity itself is an inherently system level property, as already suggested by the scaling pattern of the connections [6]. The control of such networks is of tremendous interest for problems ranging from brain-machine interface [7] to the treatment of diseases of the brain [8]. Going one scale up, it is timely to also consider networks of living organisms or even the earth's biosphere, which is estimated to consist of a few million species that are coupled together, either directly or indirectly through the environment. It is important to understand how to control such systems in order to exploit them (as in the case of commercial fishing) and also to manage wildlife, particularly in the presence of habitat loss and other perturbations. Finally, we can consider human-made systems—after all, engineering is a systems discipline by definition. Power grids, for example, are at the forefront of network research owing to the ongoing transition from analog to digital control and the smart technologies that will come with it. These networks will need to be controlled at multiple levels, and developing the required new approaches has indeed been a major drive for the current control research in network systems.

---

[1] The network systems currently investigated are also significantly different from well-mixed or well-ordered complex systems (think of a gas, a crystal) to which statistical mechanics has been traditionally applied. They are well structured, and as such cannot be characterized by few average quantities; and the equivalent to a unit cell would be essentially of the size of the entire system, which in turn leads to the issue of scale. Complexity is ultimately a statement about the dynamics (rather than the structure) of the system, which is nevertheless influenced by structure in such networks.

In this article, I offer a personal account of current challenges and recent advances in the control of complex networks.[2] I discuss in particular the unique characteristics of real complex networks from the various domains that set them apart from other systems to which control has been traditionally applied. As amply analyzed below, at the center of this longstanding interdisciplinary problem is the inherent nonlinearity of complex networks, thus making this discussion particularly appropriate for the 25[th] Anniversary issue of Chaos: An Interdisciplinary Journal of Nonlinear Science.

## II. CONTROL AND THE COMMON (EVEN UNAVOIDABLE) PROPERTIES OF REAL NETWORKS

When we look at networks coming from very different domains (intracellular, neuronal, ecological, infrastructural), it is natural to ask whether they have anything in common. Otherwise, why would we be considering several of these networks at the same time? It is widely appreciated in the network science community that these networks have several *structural* properties in common: they have a heterogeneous distribution of number of connections per node, they have community structure, they have a relatively small node-to-node distance, and so on [10]. On second thought, this is not entirely surprising because, after all, they have evolved from small to large through growth mechanisms that are somewhat related from a fundamental point of view.

My focus here is on control and, because control is a dynamical process, I want to emphasize the common *dynamical* properties that all these networks, and in fact many other real networks, tend to have:

---

[2] Thus this article focuses on the control of networks rather than "network control" as used to refer to distributed control systems that use networks of sensors and actuators to control a system that is not necessarily a network [9].

- They are governed by nonlinear dynamics, which cannot be meaningfully approximated by linear ones.
- They are dissipative dynamical systems, whose trajectories evolve toward a small portion of the entire phase space.
- They exhibit not one but multiple stable states (whether fixed points, limit cycles, or chaotic attractors).
- They are described by a large number of dynamical variables, corresponding to high phase-space dimensionality, generally of the order of the number of dynamical units.
- There are constraints on the feasible control interventions that could be applied to them, and often there are limitations to the use of feedback.
- They operate in a decentralized way, and hence tend to respond sub-optimally to perturbations with respect to any objective function they may have been designed or evolved to optimize.
- There is often noise in the dynamics and parameter uncertainty in the available models constructed from data.

Many of these properties are not yet widely appreciated in the network community. In particular, the interventions that can be implemented in trying to manipulate such systems are often of completely different nature when compared to traditional control problems. In general we are not merely limited to manipulating the variables (or parameters) by a small amount, but also there are variables that can only be manipulated in specific directions and variables that cannot be manipulated at all. For example, interventions in an ecological network are in practice often limited to suppressing (rather than increasing) the populations of specific species, while the populations of other species (e.g., endangered ones) cannot be manipulated at all. These constraints effectively limit the control to navigating a low-dimensional manifold in a very high-dimensional space, and we have to do our very best within that manifold in order to control the system. Moreover, in many cases, such as in the use of gene knockdowns to manipulate genetic networks and in the control of cascading failures (as in the approaches we have proposed and that I discuss below), there is essentially no opportunity to implement feedback. This too is a game changer, since much of traditional control theory is feedback control (and some control theorists have even defined control as the science of feedback [11]).

Also underappreciated is the fact that these systems—including most human-made, infrastructure networks—tend to respond sub-optimally to perturbations. This means that they may undergo large failures following a perturbation when in fact there are accessible states in which those large failures could be prevented. The difference between the spontaneous response of the system and the optimized response is precisely the margin that can be explored by control interventions to, for example, mitigate the propagation of a cascading failure.

This brings me to the second set of definitions that I want to discuss, if only to avoid abuse of language. First, *controllability* is a technical term [12]. It concerns the property of being able to steer a system from any given initial state to any given final state in finite time. Therefore, it is a strictly global property, and it does not say anything about our ability to keep the system at that final state. Even if we are able to drive the system from any initial state to any final state by manipulating a few degrees of freedom, if we want to keep the system at a specific final state, then, in general we will have to manipulate most dynamical variables to stabilize that state (unless the state is already stable) [13]. For being unrealistically ambitious, general results on (global) controllability are available only for linear dynamics, which, as explained above, are of limited relevance to the study of real networks (see [14-16] for attempts to extend some restricted results to nonlinear systems—also briefly discussed in the Appendix).

The term *control*, on the other hand, can refer to a number of different questions, but with emphasis on actually designing the intervention instead of just identifying when control is possible. It may concern the question of driving the system from a subset of specific initial states to a subset of final states. In particular, if you want to avoid the problem I just mentioned, you might want to limit yourself to stable final states, which is a case I will address in detail below. You may also try to stabilize states of interest, create states that do not exist, eliminate states that do exist, or any number of other actuated changes to the dynamics. All such actions would fall under the umbrella of control.

Here I will focus on control, which is the most relevant concept for the network systems I consider, but with the goal of systematically exploring all relevant possibilities. In fact, it does not even make sense in most cases to ask about the possibility of going to arbitrary states in the phase space, as that would bring the system outside the regime in which it can function. Another fundamental distinction is that network controllability studies have been centered on determining whether the system is controllable for the given control inputs (a yes/no question) and on the identification of the minimal set of driver nodes/control variables (a combinatorial optimization question on a finite set of choices). In practice, however, the hard problem that has to be solved is to determine the control signal (and the control trajectory), which involves choosing from an infinite (in fact uncountable) number of possibilities. In the rest of the article I discuss the control of network dynamics with the view of not merely identifying driver nodes but also the control signal (and hence control trajectory).

### III. WHY IS NETWORK CONTROL AN OUTSTANDING PROBLEM?

Control as a technology has been around for a couple thousand years [11]. Control theory, on the other hand is newer and has its origins often traced back to James Maxwell, who after having done most of his important work in electrodynamics, turned to what was the problem of the day: the control of the steam engine. His study of the flying ball governor established the conditions for this control system to be stable as a function of its parameters [17]. Maxwell had also studied networks: he studied networks of forces [18], which are not too different from the networks now studied in the modeling of the cytoskeleton in living cells. Maxwell himself did not combine these two topics (control and networks) but others have done so since. There are even books very explicitly focused on the control of network systems and even *complex* networks [19-22], not to mention many papers. Among the papers, I would highlight the insightful work on structural controllability published by Lin in 1974 [23], which presents an approach that takes direct advantage of the network structure. This is a paper that has been influential in recent years because network researchers are going back to those results to study linear time-invariant dynamics in networks, for the excellent reason that that case has a simple ready-to-use theory. Although not network-specific, another previous advance that is largely relevant to our discussion was the development of state-space approaches initiated in the 60s [24], which departed from the previously more common frequency-domain approaches and are in many ways a natural evolution from the work of Poincaré, Lyapunov, and others that laid the basis of modern dynamical systems theory. More broadly, control theory has a long and extremely successful history, and it is largely owing to its advances that we are now able to live in a technology-based society.

A question worth asking upfront is then whether the most fundamental problems in the control of network systems have not yet been solved. And the short answer is: *no*, they have not. Put simply, existing control approaches do not scale well with the dimension of the phase space when the realistic properties of networks discussed in Sec. II are accounted for. This severely limits their applicability to many large networks given that the size of the network is reflected in the dimension of the phase space. Naturally, there are scalable methods to address high-dimensional systems if the dynamics are linear [25], but then the dynamics of real networks are anything but linear. Likewise, there are many methods to address nonlinear dynamical systems, but in general only as long as the systems are not high dimensional. And there are methods to study nonlinear high-dimensional systems provided that the constraints on the admissible interventions are sufficiently mild [26].

Now, when we add together the various network-specific features discussed here, very quickly the problem becomes complicated and in fact untreatable by existing methods, which do not scale well enough to allow the study of network systems with hundreds, thousands, or (in some cases) millions of dynamical variables. In this context an important class of outstanding problems concern cases that involve what I call *phase-space phenomena*. Now let me distinguish those cases from scenarios that do not involve phase-space phenomena using very elementary examples.

Consider the problem of controlling the spread of an infectious disease determined by a simple epidemic spreading model when we are constrained to immunizing only a small fraction of the nodes (i.e., of individuals in the population). A naïve approach would consist of selecting nodes at random and immunizing them. An improved approach would consist of immunizing high-degree nodes, which can be done efficiently by immunizing random network neighbors of randomly selected nodes [27]. Applicable to social networks for which no global information is available, this strategy draws from the general network property that in degree-heterogeneous networks the average number of neighbors of neighbors is always larger than the average number of neighbors [28], the difference being larger the larger the heterogeneity (the same property can in fact be exploited for the early detection of epidemic outbreaks [29]). What is interesting about this example is that everything I described is entirely based on local information and entirely determined by the structure of the network. At that level, and leaving aside the question of whether real epidemiological processes can be reduced to such a simple description, one can say that the problem is solved without any reference to the specifics of the dynamics or of processes that take place in a phase space. This is the case because the control intervention is already defined (it consists of immunizing the node) and the problem reduces to identifying the nodes to intervene on.

To proceed, let us take a very simple system that can be regarded as a network and that exhibits phase-space phenomena: $\dot{x}_1 = x_1 + u_1(t), \dot{x}_2 = x_1$, where $u_1(t)$ is the control and the other terms define the 2-dimensional autonomous dynamics of a two-node network [30]. Kalman established the condition for systems of the form $\dot{x} = Ax + Bu(t)$ of any dimension $n$ to be controllable [12]. The condition is that the matrix $K = [B\ AB \cdots A^{n-1}B]$ be full rank, which is easy to test—unless the system is really high dimensional, but even in those cases there are computationally efficient methods that can be used to test this condition [31]. (Note that, akin to testing this condition for a given matrix $B$ is the problem of identifying the minimal set of driver nodes or control inputs for the system to be controllable, which corresponds to identifying an "optimal" matrix $B$ satisfying the Kalman rank condition for $K$—another problem for which computationally efficient algorithms have been developed over the years [32] and which has received recent attention following the publication of an elegant new algorithm in [33].) Therefore, the problem of determining whether a linear system is controllable is essentially a solved problem whose answer is, as in the epidemic example above, entirely determined by the network structure. In particular, it is clear that the 2-dimensional system above is controllable by means of an input signal $u_1(t)$.

Now suppose that, instead of merely checking whether the system is controllable, we are asked to build the control signal $u_1(t)$ or, equivalently, the control trajectory. As it will become clear soon, this problem is far more involved, since it involves phase-space phenomena. To appreciate that, consider the phase space of the system above shown in Fig. 1, where the arrows in the background represent the vector field of the autonomous portion of the system, and assume that our goal is to steer the trajectory from the open symbol to one of the solid symbols. For a control of the given form, which only actuates the first of the two variables, this can be achieved but requires crossing to the left past the dashed line at $x_1 = 0$, to use the autonomous flow itself to steer the trajectory downwards before the control of $x_1$ can move it toward the target point on the right side. This is the case no matter how close this final state is to the initial one.

Therefore it follows from this that the control trajectories are *nonlocal*. This property—which was established and analyzed in detail in [30]—has many implications, which are not yet fully appreciated. One of them is that, while this is a simple situation in a 2-dimensional system, in a large network in which the number of control inputs $q$ is much smaller than the number $n$ of nodes in the network—a scenario pursued by seeking to identify a minimal set of driver nodes—the system becomes numerically uncontrollable even when it satisfies the Kalman rank condition. Why? Because the Gramian, a matrix that we have to effectively invert in order to calculate the control trajectory, becomes ill-conditioned and hence numerically singular. This is so because in such a case there are many hyperplanes analogous to the dashed line in the 2-dimensional example (Fig. 1) that the control trajectory has to cross in order to reach the final state while being actuated by that small number of control inputs. As the dimension of the system goes up (more precisely, as $n - q$ increases), the length of the control trajectory increases very rapidly. The longer the control trajectory the larger the condition number of the Gramian matrix. The condition number of the Gramian grows exponentially as the number of control inputs reduces or the number of degrees of freedom increases. Therefore, this is not a problem simple to solve and it is one that, like sensitive dependence on initial conditions in the study of deterministic chaos, cannot be avoided by just increasing the precision of the calculations.

As proposed in [30], we need a controllability criterion that accounts not only for the existence but also for the actual computability of the control interventions: the system is controllable in practice if and only if the controllability Gramian has full *numerical rank*. The numerical rank can be interpreted as the number of singular values that are larger than a predefined numerical threshold, and as such involves a criterion for deciding when a number should be treated as zero given the available precision of the numerical computations. This criterion, which should not be confused with continuous indexes based on condition numbers [34], also has the great advantage of increasing modeling robustness.[3]

The moral of the story is that if our goal is just to determine whether the system is controllable and what the minimal set of driver nodes would be, the problem is analogous to the epidemic problem discussed earlier—that is, we avoid the phase space altogether and the problem is in principle manageable. Now, as soon as we try to actually control the system, which requires determining the control trajectories numerically, the problem becomes much more complicated. Therefore, if the question concerns actual control (not just controllability), even linear systems can be difficult to handle when they are sufficiently high dimensional. And sufficiently high dimensional here means just a few hundred degrees of freedom [30].

Another implication of the nonlocality of the control trajectories, which is subtler but extremely important, is that around typical points we cannot linearize a nonlinear system to then use the results established for the linear system to control the nonlinear one. Why? Because nonlocal control trajectories can go outside the region in which linearization is valid, and the results will be inconsistent even if the target states are limited to be very close to the initial ones (as discussed in some detail in Ref. [13]). Therefore, while there are other procedures in which linear methods can be used in the control of nonlinear systems (as in the case of first-order controllability around an equilibrium point [59]), this is not one of them. It follows as a corollary of this no-go result that the control of a network with hypothetical linear dynamics is not informative of the control of the actual dynamics even if the network structure is the same. In particular, the former being controllable is neither sufficient nor necessary for the latter to be controllable, as demonstrated for instance in [13] through both theory and examples.

In the context of nonlinear systems, traditional approaches of control have emphasized scenarios in which the system is close to normal operating conditions. While these scenarios remain extremely important, in the study of networks it is also important to consider scenarios in which the system is far from the desired state (and/or the opportunities to benefit from feedback are limited, such as in the case of cellular reprogramming or in the control of a propagating cascade and other adverse conditions).

---

[3] For example, consider two systems with the same matrices $A$ and $B$, except that in the first system the nodes have self-dynamics (i.e., the diagonal elements of $A$ are nonzero) while in the second system there is no self-dynamics (i.e., $A$ has null diagonal). Theoretically, the first system requires multiple control inputs [33] while the second can be controlled by a single control input [35]. But that is only true at arbitrarily large precision. Numerically this otherwise surprising difference disappears, and both systems will generally require a large number of control inputs to be controlled [36].

Our emphasis here will therefore be on the control of network dynamics far from equilibrium, which remain largely underexplored and is an extremely timely area of research, with tremendous potential for new developments in both theory and applications.

## IV. SOLUTION OF A NONLINEAR NETWORK CONTROL PROBLEM

The control of networks with nonlinear dynamics is significantly more difficult than of those with linear dynamics (see also the Appendix). Given the difficulties identified above already present in the case of linear dynamics, one may wonder whether we can establish any general control method that would work in the case of nonlinear dynamics. The short answer is *yes* and for a range of conditions, but to start—and illustrate the essence of the problem—I will consider the most favorable scenario: network systems that has purely deterministic equations of motion (which I will assume to be ordinary differential equations, though that is not essential) and that have no parameter uncertainty, delays, or other complications. Later we can lift these assumptions.

Let's assume that the problem we want to solve is the one in which a cascading failure has been triggered and our goal is to mitigate its propagation. Or the closely related problem in which no cascade has been triggered and our goal is to repurpose the network from its original function to a new one (as in the process of cell reprogramming). In either case the problem can be formally interpreted as one in which the trajectory of the system is away from the desired attractor and would autonomously evolve to (or be at) a different (undesirable) attractor. Our task is to design a control intervention to address the situation. That is, an intervention that would bring the system to the basin of attraction of the desired attractor, from where it can evolve autonomously to the desired attractor itself.[4] In the case of a cascading failure, the problem is particularly interesting because it sets upfront limits on the time we have available to do calculations for the control decisions and to implement them. In the graph representation of the system—where the constraints on feasible interventions often limit the control perturbation to a small set of nodes—what we would be trying to do is essentially to trigger a compensatory cascade that would neutralize the original one, which is in course. That seems an impossible task, but mainly because the graph is not the most insightful representation of the problem (even if, as it is usually the case, it is the representation in which the consequences of cascading failures are observed). Like any dynamical problem, this one too is best understood in the phase-space representation, which I adopt to discuss the solution established in Ref. [39].

Given the assumed conditions, we can describe the dynamics through equations of the form $\dot{x} = F(x; \beta)$, where the most important thing to note is that the system has dynamical variables, denoted by vector $x$, and parameters, denoted by vector $\boldsymbol{\beta}$—both high-dimensional vectors. The network structure is accounted for by $F$. Suppose we are focusing on attractor $A$, which has a basin of attraction $\Omega_{\boldsymbol{\beta}}(A)$, and that at time $t_0$ the system is at a state $x_0 \notin \Omega_{\boldsymbol{\beta}}(A)$. If our control is based on actuating the dynamical variables, the task is then reduced to designing a control perturbation $\Delta x_0^A$ that could bring the system to a new state $x_0' = x_0 + \Delta x_0^A \in \Omega_{\boldsymbol{\beta}}(A)$ (expressed in terms of an impulse perturbation just to keep the logic simple—this too is not essential). Put as such the problem is trivial: all we have to do is to take $\Delta x_0^A$ as the vector difference between the current state $x_0$ and a point at the desired attractor $A$. But reality is not so simple: there are usually constraints on the admissible interventions. In a biochemical network, for example, even though individual reactions can be manipulated in both directions, due to bottlenecks it is usually easier to down-express a pathway than to over-express it. In an ecological network, it is usually easier to suppress a species population than to increase it. For this reason, we have to account for inequality constraints of the form $\boldsymbol{g}^x(x_0', x_0) \leq \boldsymbol{0}$ (with the convention that the inequality applies to each component). We also have to account for equality constraints, $\boldsymbol{h}^x(x_0', x_0) = \boldsymbol{0}$, which often represent variables that cannot be actuated, such as nodes that are not accessible to manipulation or whose manipulation could lead to adverse effects; in the examples just given these could be essential biochemical reactions or endangered species. With these constraints the problem becomes highly nontrivial, even under the seemingly favorable conditions assumed this far.

---

[4] Note that the challenge is to steer the trajectory toward a specific different attractor (hence, across different ergodic regions of the phase space). It is not important whether the attractor is a fixed point—it can be periodic or even chaotic since, once the attractor is reached, simple methods can be used to manipulate that dynamics within it [37, 38].

The main reason this problem is difficult is because, on the one hand, these constraints generally prohibit bringing the system directly to the desired attractor $A$. The problem can still be solved if we identify an intervention that instead brings the system the basin of attraction $\Omega_{\boldsymbol{\beta}}(A)$. But then, on the other hand, there is no general analytical or numerical method to locate basins of attraction in high-dimensional phase spaces (analytical methods, such as those based on Lyapunov functions, generally offer conservative estimates and numerical methods suffer from lack of scalability). As illustrated in Fig. 2, the real problem in the control of such nonlinear networks reduces to trying to reach the intersection between the region of feasible interventions defined by the constraints and the desired basin of attraction, which is not known. Upfront, we can't even tell whether there is an intersection and hence whether the problem has a solution. What we do know is, of course, the dynamics as defined by the equations of motion. Using this local piece of information we have been able to solve this puzzling problem, which is global in nature, by establishing a method that effectively brings the state of the system to (this intersection with) the basin of the desired attractor even though we do not know explicitly where the basin of attraction is [39].[5]

Before discussing this solution, note that instead of bringing the state to the attraction basin we could have sought to bring the attraction basin to the state, as illustrated in Fig. 3. This could be achieved by actuating the parameters instead of the dynamical variables (in an ecological network this would correspond to manipulating growth or mortality rates instead of population abundances). Specifically, we would seek to change the parameters from $\boldsymbol{\beta}$ to $\boldsymbol{\beta}'$ such that $x_0 \in \Omega_{\boldsymbol{\beta}'}(A')$, where $A'$ a smoothly deformed version of the desired attractor for the modified parameters (assume, for simplicity, that this involves no bifurcations). Here too we would have to respect inequality and equality constraints, $g^\beta(\beta',\beta) \leq 0$ and $h^\beta(\beta',\beta) = 0$, which in the ecological example could mean that growth (mortality) rate can only be reduced (increased) and some species cannot be manipulated. Once in the basin $\Omega_{\beta'}(A')$ the trajectory would autonomously approach attractor $A'$. Once it is close to $A'$ we could then slowly relax the parameter perturbation, so as to cause the trajectory to follow the attractor until it changes back to $A$—the desired state.

---

[5] It goes without saying that the intersection of the feasible region with the target basin of attraction will generally depend on time of the intervention. A dramatic illustration of this dependence was shown in Ref. [40], where examples were given of extinction cascades in food-web networks that could be prevented entirely by the suppression of species that would otherwise be eventually extinct by the cascade.

The method we have developed to reach the target basin of attraction involves two main elements [39]. First, that we know the desired attractor (which is in general much easier to determine than its basin) and that the dynamical equations allow us to forecast the future trajectory. Second, that a finite-size control perturbation that could bring the system to the target basin can be built by iteratively calculating small perturbations using information provided by the (local) equations of motion. Specifically, on the forecast trajectory we can locate the closest approach point to the desired attractor and then ask: "in what direction should we change the state $x_0$ at time $t_0$, by a perturbation no larger than $\varepsilon$ while respecting the given constraints, so that the closest approach point of the new trajectory will be closer to (and at the smallest possible distance of) the desired attractor?" We have implemented a computationally efficient algorithm to address this question, which uses the variational equation to map perturbations forecast at the closest approach point to the optimal one at the initial time $t_0$. By repeating this process, the closest approach point will successively approach the desired attractor and, if a solution can be found, it will eventually converge to the attractor, at which point the control perturbation has crossed into the basin of attraction (even though no explicit information about its location was used). The result is a gradient descent-like method, which—in contrast with ordinary gradient descent optimization schemes—is in this case tailored to solve a problem that is generally nonconvex due to the constraints. A ready-to-use version of this algorithm is available through Ref. [41].

The method is highly scalable and effective. The computational cost scales with the number of dynamical variables as $n^{2.5}$, and it can therefore address very large networks in a computationally inexpensive manner.[6] Its effectiveness has been demonstrated in applications both to real networks from various domains and to ensembles of model networks. Realistic applications of this approach, and its variants [40], have included the control of de-synchronization instabilities in power-grid networks, identification of interventions to mitigate extinction cascades in food-web networks, and the identification of therapeutic interventions for an epigenetic form of cancer [39].

---

[6] Incidentally, here is where we take advantage of the sparsity that sets networks apart from other dynamical systems. The approach can be applied to any dynamical system at the estimated computational cost of $O(n^{3.5})$, but for networks this cost is reduced by a full power of $n$ provided that the number of variables is approximately proportional to the number of nodes and that the average degree remains essentially constant, as is the case in many network models.

A relevant question is whether the method can cross intermediate basins of attraction. The answer is *yes,* because the process resets itself every time it crosses a separatrix. And performance is not adversely affected in systems with complex or fractal basins of attraction—its effectiveness has been demonstrated even for riddled basins of attraction [39]. The core method also has the merit of being easily adaptable to address more general dynamics. For example, if instead of taking the first control intervention that crosses into the target basin of attraction we add an additional (rather straightforward) optimization step to minimize (under the given constraints) the time it takes to reach the neighborhood of the attractor, the resulting method is also effective in the presence of moderate noise and parameter uncertainty; the attractors and basins of attractions are in this case defined with respect to the deterministic, well-defined portion of the dynamics. Finally, even when it is impossible to reach the desired attractor (e.g., if the constraints are too restrictive), in practice this method will tend to bring the system to an attractor whose properties of interest are more similar to those of the desired one, as shown in Fig. 4 for an associative memory network. This is expected, in particular, when such properties depend continuously on the dynamical variables (and hence on the location of the attractors), as it is often the case.

Is there a relation between the identified control intervention and the structure of the network for general networks with nonlinear dynamics? Yes, there is, if the network elements (nodes and edges) are all comparable, so that the system is dominantly determined by network structural parameters. In particular, there will be a correlation between the likelihood that a node will belong to a successful control set and its centrality measures in the network—such as its degree in the case of random networks. But this is generally not true in real networks since real networks are defined by many structural *and* dynamical parameters, and the topological parameters associated with the structure of the network are just part of them. It is therefore essential to consider the problem in the phase space. After the calculations are done in the phase space, it is instructive to go back to the graph representation to interpret the result. This will often lead to insightful conclusions about the role of the network structure (as shown in some detail in Ref. [40] for food-web networks). But this is a posteriori analysis; it's not something that we can use to solve the control problem by inspection of the network structure.

## V. OUTLOOK ON CURRENT AND FUTURE RESEARCH

The approach discussed in Sec. IV to control nonlinear network dynamics can be generalized in many ways. For example, the effectiveness of an intervention will generally depend on the time at which it is implemented. It is therefore advantageous to consider interventions not only at the initial time but also at any later time. Moreover, instead of relying on impulse control we can consider continuous interventions formulated, for example, in terms of model predictive control. Another generalization would be to consider closed-loop control for network problems that can benefit from real-time feedback and for which feedback can actually be implemented. Such generalizations would be useful in practice, and the quality of the results would only improve. In view of practical applications, it would also be useful to consider situations in which the state of the system is only partially known.

Another scenario extremely important to consider is the following. I have discussed robustness to small noise—a case that can be addressed by the same framework considered above [39]—but now suppose that noise is large enough not only to make the basin boundaries fuzzy but also to induce transitions between basins of attraction. In this case, which is common in biological networks [49, 50], a question of interest is whether we can control the response of the system to noise in order to induce and/or inhibit specific transitions and ultimately control the occupancy of the stable states by manipulating tunable parameters in the system. An elegant solution to this problem was recently presented in [42], where the following elements were combined into a very effective and efficient control approach: 1) large deviation theory was used identify the transition paths as the least action paths and to calculate the corresponding transition rates; 2) the transition dynamics of the original network were reduced to a Markov process on a network of state transitions between the attractors, on which control was implemented. This way, by using a new network to solve the original network problem, we effectively reduce a high-dimensional problem to a sequence of one-dimensional ones, resulting in a scalable approach. This approach changes the system response to noise instead of the noise itself by suitably modifying the underlying quasi-potential, and is therefore analogous to the approach recently undertaken to design mechanical material networks with unusual phase transitions by manipulating the underlying free energy function [43].

The key common property of the nonlinear-dynamics control approaches reviewed here [39-42] is that they all explore the structure of the phase space. Instead of merely optimizing an objective function, they incorporate information about the attractors and take advantage of the fact that the phase space is patched into basins of attraction (even if their location is not explicitly known). The advantages are two-fold: first, this makes the methods more robust and scalable, as it suffices to reach the attraction basin (a full-dimension set of the phase space) rather than a point of the attractor (a lower dimensional set); second, this makes the methods more effective, as it saves us from the scenario in which the phase-space point optimizing the function of interest happens to be not only outside all attractors but also outside the basin of the most desirable attractor (in this case, the system would evolve back to an undesirable state as soon as the control is switched off). A different context in which the structure of the phase space has been exploited was in the control of conservative systems [44, 65], where it is beneficial to consider the partition of the phase space into ergodic components. Aside from the numerical advantages, the analysis of the structure of the phase space facilitates conceptual understanding of the problem, as previously demonstrated in the qualitative study of differential equations.

Finally, I should note that there are several other currently active lines of research concerning the control of network systems, which are pursued by various communities and which do not necessarily involve the assumptions or scenarios that I invoked here. Within the network community, significant part of the attention has been devoted to the study of controllability and observability, particularly of linear time-invariant systems [33, 35, 30] (but see also [51, 52, 45] for nonlinear cases). Attention has also been given to the study of control processes in consciously simplified models, particularly to guide the formulation of hypotheses for processes away from equilibrium [57, 58]. On the other hand, pinning control [56] has received significant attention within the nonlinear dynamics community in connection with network synchronization (a topic that has attracted increasing interest [53-55]) and consensus processes, where the control strategy is based on introducing a leader to directly influence the dynamics of a selection of nodes [46-48]. In the control community, decentralized and distributed control [60, 61, 20], which concern systems composed of a large number of interacting subsystems (including networks), are classic areas that continues to be extremely active, particularly in the context of large-scale systems in various contexts [62-64, 9]. The control of flocking, schooling, moving sensors, and collective behavior in general in networks of autonomous agents is a related line of research that has received significant recent attention. Control in the context of transportation networks, supply chains and operations research in general is yet another very active field of current research. There are also new applications that are stimulating significant research of which I would highlight the ongoing development of autonomous automobiles and smart grids—both involving stimulating network control questions.

**Appendix: Nonlinear Control and Controllability**

Nonlinear systems are significantly different from linear ones but some general results can still be established. For example, a system of the form $\dot{x} = f(x, u)$, where $u$ is the control, will be (globally) controllable if for every $x$ the set of realizable vectors $f(x, u)$ contains $0$ in its interior [66]; this is, of course, a sufficient but not necessary condition (to which even the simple system in Fig. 1 serves as an example). A closer extension of the Kalman rank condition can be established by defining a suitable nonlinear controllability matrix in terms of Lie brackets [67]—an operator that given two vector fields defines another vector field that essentially measures the non-commutativeness of the corresponding flows, and hence provides information about the directions that can be achieved by combining those two fields. The controllability is again associated with a full rank condition, but with two important differences from the linear case: the condition is in this case only necessary and the controllability it speaks to is local [16, 67]. For this reason, in this context it is often more useful to consider other properties, such as reachability and accessibility, for which stronger (necessary and sufficient) conditions have been established [15, 67].

**ACKNOWLEDGEMENTS**

The author thanks Aleksandar Haber and Jie Sun for informative discussions. This work is funded by the National Science Foundation under grant DMS-1057128, a Multidisciplinary University Research Initiative under grant ARO-W911NF-14-1-0359, and the National Institutes of Health under grant NIGMS-1R01GM113238.

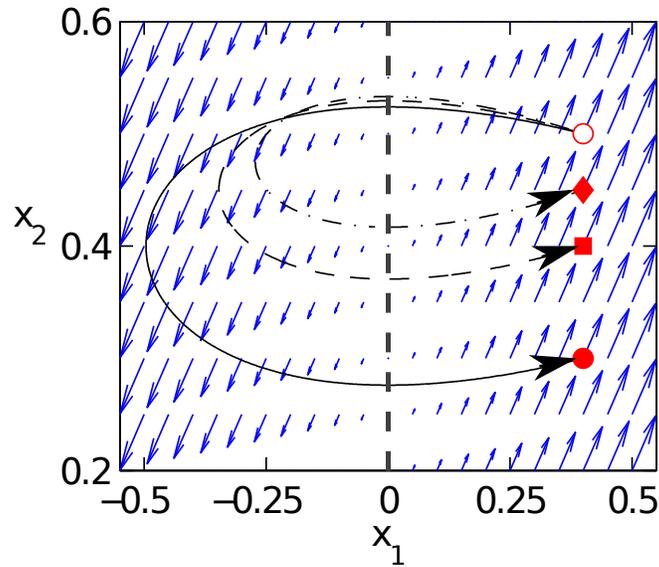

**Figure 1.** Example of a system that is controllable but whose control trajectories are nonlocal (adapted from Ref. [30]).

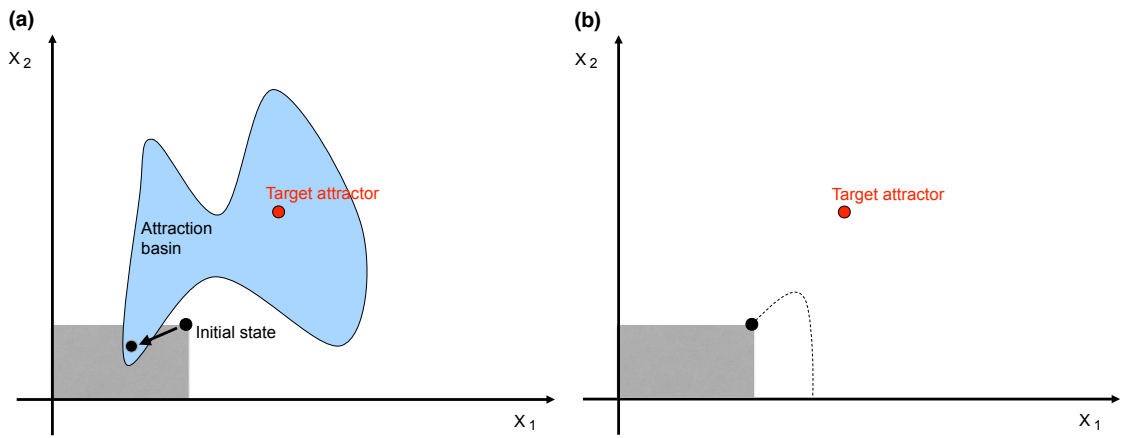

**Figure 2.** Illustration of a control problem in which the variables are constrained not to be increased by the intervention, which prohibits steering the trajectory directly to the desired attractor. (a) A solution, which consists of bringing the system to the basin of the desired attractor. (b) The problem as it appears to the observer when the basin of attraction is not known, as in the case of networks with high-dimensional phase spaces. The dotted line indicates the future evolution of the uncontrolled system.

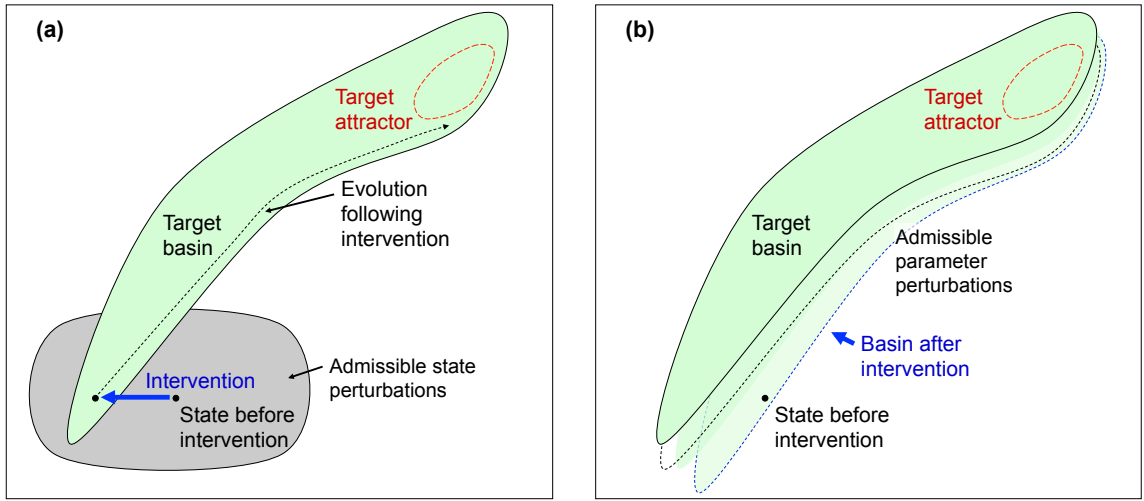

**Figure 3.** Schematic illustration of the control approach, for interventions based on manipulating (a) dynamical variables and (b) system parameters.

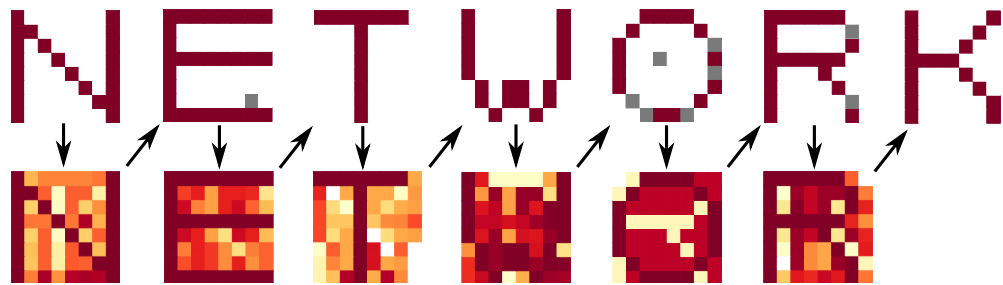

**Figure 4.** Example of associative memory network in which each letter of the word "NETWORK" is stored as an attractor. The network is 8x8 and each node is a phase oscillator color-coded by the phase; in stationary states each node can be in one of two states: in phase or anti-phase with respect to a reference node (marked as ON and OFF pixels, respectively). In this illustration the control problem is to drive the network from the attractor representing a letter to the attractor presenting the next letter by only manipulating OFF-pixel nodes. The control interventions are indicated by the vertical arrows, and the subsequent evolutions toward the attractors are indicated by the oblique arrows. The gray pixels mark errors, which means that in some cases the system converged to a different, parasite attractor. The reached attractors are, nevertheless, remarkably similar to the intended ones, indicating that the method is robust even when the desired solution is not possible. (Figure adapted from Ref. [39]).